\documentclass[aps,prb,twocolumn,superscriptaddress,showpacs,floatfix,amsmath]{revtex4}

\usepackage{tipa}
\usepackage{bbding}
\usepackage{txfonts}
\usepackage{amssymb}
\usepackage{graphicx}
\usepackage[T1]{fontenc}
\usepackage[colorlinks,
linkcolor=blue, urlcolor=blue, anchorcolor=blue, citecolor=blue]{hyperref}
\begin{document}

%\preprint{HEP/123-qed}

\title{Realisation of adiabatic and diabatic CZ gates in superconducting qubits coupled with a tunable coupler}

\author{Huikai Xu}
\thanks{H. Xu and W. Liu contributed equally to this work.}
\affiliation{Beijing Academy of Quantum Information Sciences, Beijing 100193, China}
\author{Weiyang Liu}
%\newcommand
%\collaboration
\thanks{H. Xu and W. Liu contributed equally to this work.}
\affiliation{Shenzhen Insititute for Quantum Science and Engineering, Southern University of Science and Technology, Shenzhen 518055, China}

\author{Zhiyuan Li}
\affiliation{Beijing Academy of Quantum Information Sciences, Beijing 100193, China}

\author{Jiaxiu Han}
\affiliation{Beijing Academy of Quantum Information Sciences, Beijing 100193, China}

\author{Jingning Zhang}
\affiliation{Beijing Academy of Quantum Information Sciences, Beijing 100193, China}

\author{Kehuan Linghu}
\affiliation{Beijing Academy of Quantum Information Sciences, Beijing 100193, China}

\author{Yongchao Li}
\affiliation{Beijing Academy of Quantum Information Sciences, Beijing 100193, China}

\author{Mo Chen}
\affiliation{Beijing Academy of Quantum Information Sciences, Beijing 100193, China}

\author{Zhen Yang}
\affiliation{Beijing Academy of Quantum Information Sciences, Beijing 100193, China}

\author{Junhua Wang}
\affiliation{Beijing Academy of Quantum Information Sciences, Beijing 100193, China}

\author{Teng Ma}
\affiliation{Beijing Academy of Quantum Information Sciences, Beijing 100193, China}
\author{Guangming Xue}
\email{Corresponding author: xuegm@baqis.ac.cn}
\affiliation{Beijing Academy of Quantum Information Sciences, Beijing 100193, China}

\author{Yirong Jin }
\email{Corresponding author: jinyr@baqis.ac.cn}

\affiliation{Beijing Academy of Quantum Information Sciences, Beijing 100193, China}
\author{Haifeng Yu}
\affiliation{Beijing Academy of Quantum Information Sciences, Beijing 100193, China}
%\date{\today}

\begin{abstract}

High fidelity two-qubit gates are fundamental for scaling up the superconducting number. 
%In general, the static ZZ interaction between two qubits has a serious impact on the gate fidelity. Therefore, 
We use two qubits coupled via a frequency-tunable coupler which can adjust the coupling strength, and demonstrate the CZ gate using two different schemes, adiabatic and diabatic methods. The Clifford based Randomized Benchmarking (RB) method is used to assess and optimize the CZ gate fidelity. The fidelity of adiabatic and diabatic CZ gates are $99.53(8)\%$ and $98.72(2)\%$, respectively. We also analyze the errors induced by the decoherence. 
%, which are $92\%$ of total for adiabatic CZ gate and $46\%$ of total for diabatic CZ gates. The adiabatic scheme is robust against the operation error. But the diabatic scheme is sensitive to the purity and operation errors.
 Comparing to $30$ ns duration time of adiabatic CZ gate, the duration time of diabatic CZ gate is $19$ ns, revealing lower incoherence error rate $r'_{\rm{incoherent, int}} = 0.0197(5)$ than $r_{\rm{incoherent, int}} = 0.0223(3)$.    
\end{abstract}


\maketitle

%\section{Introduction}
A programmable superconducting information processor which consists of a two-dimensional array of 53 transmon qubits has been demonstrated to achieve the supremacy for a specific computational task\cite{Arut19}. A fast, high-fidelity gate scheme is the key to reaching this milestone. %It's a milestone for engineering fault-tolerant logical qubits.
%High-fidelity of two qubits gates are critical to build large scaling up quantum information processors.
For superconducting transmon/Xmon qubits\cite{koch07,Bare13}, there are a variety of proposals to realize two-qubits gates which can be divided into three main classes. The first class is implemented with frequency-tunable qubits with the static couplings. 
Interactions between qubits can be turned on and off by tuning the frequency of qubits. % to realize the two qubits gates\cite{sch03}. 
In particular, by tuning the qubits to make the $|11\rangle$ state in resonance with $|02\rangle$ state, the controlled-Z (CZ) gates can be realized\cite{sch03,Stra03,dic09,Yama10,bare14}. Furthermore, parametrically modulated qubits make particular states in resonance to realize the $i$SWAP gate and the CZ gate\cite{cald18}. In such schemes, each qubit need an additional magnetic flux bias line, which make the circuit complicated when scale up qubits numbers. The second class is implemented with frequency-fixed qubits and the static coupling strengths\cite{rige05,leek09,Hutc17}. The qubits gates are realized with all microwave driver methods, such as the cross-resonance (CR) gate\cite{rige10,chow11,chow13,shel16,pole12,Groot10,Groot12}. This scheme is facing frequency crowding problems limiting the circuit integration. The last class is implemented with frequency-fixed or frequency-tunable qubits coupled with the additional tunable coupler. The general scheme is that two frequency-fixed qubits have a large frequency detuning to eliminate the ZZ interaction which can induce the single qubit gate errors, then parametrically modulated coupler method are used to realize the $i$SWAP and CZ gates\cite{ganz20}. A more advantage scheme is to treat the tunable coupler as a switch which can quickly turn on/off the interactions between adjacent qubits\cite{Ashhab06,Ashhab07,chen14,yan18,xu20,liX20,han20,Coll20,McKay19}. The coupler will turn off the interaction between adjacent qubits when the single qubit gate is implemented. In contrast, two-qubit gates are activated when the coupler turns on the interactions. Another advantage is that qubits do not need large frequency detuning and the frequency crowding problem can also be alleviated\cite{Arut19}. %So, qubits coupled with a switch is a more advantage design.

By now, the CZ gate fidelity is as high as 99.7\%\cite{Kyaw20}. Fast gate and low operation errors are two keys to improve the two-qubit gates fidelity. In general, long gate time means more incoherence errors. However, shorter gate time will cause more operation errors. To solve this conflict problem, a fast adiabatic protocol is proposed\cite{bare14,moll18,rol19} and the gate time is about 30 $\sim$ 40 ns. Then, the nonadiabatic gates\cite{li2019,bare19,Foxen2020} are found to have fast gate times, which can eliminate incoherent errors.%The process of adiabatic gates can be described as moveing the state $|11\rangle$ close into the anti-crossing energy points between. In a small gate time, optimal the gate waveforms make the gate errors keeping at low levels. can be understanded like this. And the fast adiabatic CZ gate.
%In the present work, we investigate adiabatic and non-adiabatic CZ gates in two Xmon qubit coupled with a tunable coupler switch systems. In Section II we will briefly introduce the description of the system Hamiltonian. Section III introduce the experiments
%\section{Experiment}

In the present work, we investigate adiabatic and diabatic CZ gates in two Xmon qubits coupled with a tunable coupler system, as shown schematically in Fig.~1(a) and (b). Each qubit is capacitively coupled with a coupler which can be seen as a Xmon qubit. Two qubits are also directly coupled via the capacitance. The Hamiltonian of system can be written as \cite{yan18}:
\begin{equation}
\begin{split}
H/\hbar =
& \sum_{i=1,2,c}\omega_{i}a_{i}^{+}a_{i} +\frac{\alpha_{i}}{2}a_{i}^{+}a_{i}^{+}a_{i}a_{i} \\
& +\sum_{i \neq j}g_{ij}(a_{i}^{+}a_{j}+a_{i}a_{j}^{+}),
\end{split}
\end{equation}
where $\omega _{i}\left(i = 1,2,c\right)$ are the frequencies of each qubits ($Q_1$, $Q_2$) and the coupler ($C$), $a_i^{+}$ and $a_i \left(i = 1,2, c\right)$ are creation and annihilation operators of $Q_1$, $Q_2$ and $C$, respectively. $g_{1c}(g_{2c})$ is the coupling strength between $C$ and $Q_1(Q_2)$ and $g_{12}$ is the direct coupling strength between $Q_1$ and $Q_2$. The maximum frequencies of $Q_1$, $Q_2$ and $C$ are $\omega_{1}^{\rm{max}}/2\pi = 4.508$ \rm{GHz}, $\omega_{2}^{\rm{max}}/2\pi = 4.701$ \rm{GHz} and $\omega_{c}^{\rm{max}}/2\pi = 5.419$ \rm{GHz} and the anharmonicities are $\alpha_{1}/2\pi = -290$ MHz, $\alpha_{2}/2\pi = -306$ MHz, and $\alpha_{c}/2\pi = -124$ MHz, respectively. The effective coupling strength\cite{yan18,koch07} between $Q_1$ and $Q_2$ is
%To well understand how the coupler to adjust the coupling strength between $Q_1$ and $Q_2$, the Schrieffer-Wolff transformation
\begin{figure}[t]
	\includegraphics[width=0.48\textwidth]{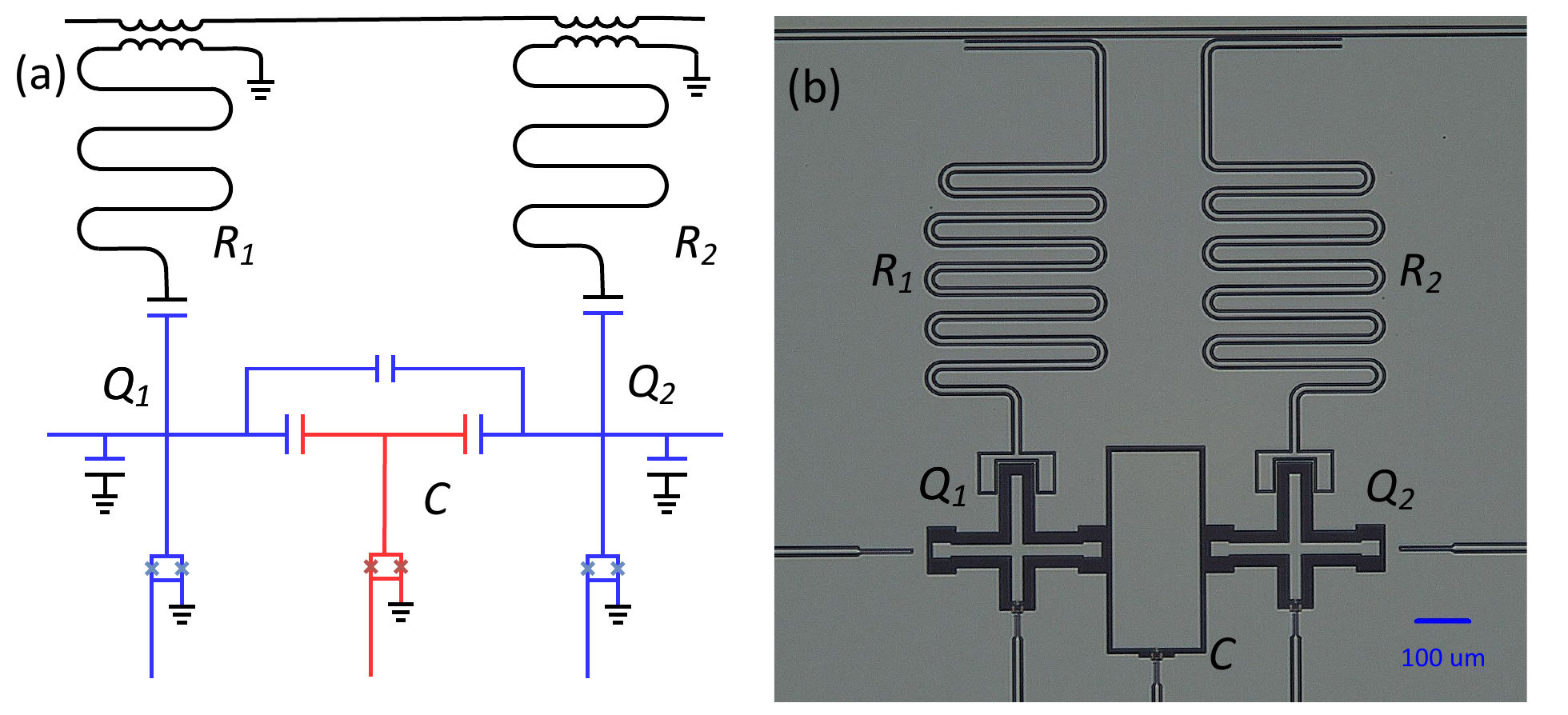}
	\caption{(a) and (b) are the schematic electrical circuit and optical micrograph of three Xmon qubits($Q_1$, $Q_2$ and $C$). $Q_1$ and $Q_2$ are used as the computational qubits with the $XY$ and $Z$ control and coupled with the $\lambda/4$ resonators $R_1$ and $R_2$ for quantum state readout. Qubit $C$ can be seen as the tunable coupler($C$) with only the $Z$ control. 
		%The coupling strength between $Q_1$(or $Q_2$) and $C$ is about $100$ MHz. The directly coupling strength between $Q_1$ and $Q_2$  is $5$ MHz.
	%(c) Schematic of pulse sequence to measure the coupler $C$ spectrum. Driven coulper $C$ through qubit $Q_1$ XY control line while biased the coupler by a square pulse. Then readout the excited population of qubit $Q_2$ after A $X_{\pi}$ pulse on qubit $Q_2$. (d) Measured the spectrum of coupler $C$ coupled with qubits $Q_1$ and $Q_2$.
	%(e) Schematic of 
	$i$SWAP experiment. The $Q_1$ is excited, then tune $Q_2$ resonate with $Q_1$. Biasing the flux of $C$ can change the interaction strength between $Q_1$ and $Q_1$. (f) is the $i$SWAP oscillation versus flux bias of $C$. (g) Fourier transition of $i$SWAP oscillation in (f). The light line indicate the total coupling strength. $\tilde{g}$ can be adjusted from 0.40 to 40 MHz.
}
\end{figure}
\begin{equation}
\tilde{g} \approx \frac{g_{1c}g_{2c}}{2} \left(\frac{1}{\Delta_{1c}}+\frac{1}{\Delta_{2c}}\right)+g_{12},
\end{equation}
%$\tilde{g}$ can be seen as the effective coupling strength between $Q_1$ and $Q_2$. 
where $\Delta_{ic} = \omega_{i}-\omega_{c} \left(i = 1,2,c\right)$.
In experiments, the value of $g_{1c}\left(g_{2c}\right)$ can be extracted from the spectrum of qubits or coupler.Figure 2(a) is the spectrum of coupler $C$. Since coupler $C$ does not have the readout cavity, we use a novel method indicated in the inset of Fig. 2(a). The coupler $C$ could be driven by a microwave pulse through the XY control line of $Q_{1}$. If the microwave resonates with coupler $C$ which is excited from $|0\rangle$ to $|1\rangle$, the frequency of $Q_2$ will shift due to Lamb shift. Therefore, applying a $\pi$-pulse on qubit $Q_{2}$, the population of excited state $Q_{2}$ will be decreased. The frequency of the coupler $C$ coupled with qubits $Q_1$ and $Q_2$ versus the amplitude of the bias pulse ($V_b$) of the coupler $C$, as shown in Figa. 2(a). The anti-crossing shows that the coupling strength $g_{1c}/2\pi\left(g_{2c}/2\pi\right)$ is $100$ MHz. The direct coupling strength value $g_{12}$ can be derived from $\tilde{g}$ which can be extracted from $i$SWAP experiment. As indicated by the inset of Fig. 2(c), setting up the frequencies of $Q_{1}$ and $Q_2$ on resonance ($4.110$ GHz) and then sweeping the amplitude $V_b$ and duration time $\tau$ of the bias pulse, a chevron pattern of $i$SWAP experiments can be measured, as shown in Fig. 2(b). The spectrum of two-qubits coupling strengths can be obtained by the Fourier transform of $i$SWAP time traces as shown in Fig. 2(c) and the light line in Fig. 2(c) indicates the coupling strength $\tilde{g}/2\pi$ which can be changed from 0.4 \rm{MHz} to 80 \rm{MHz} by adjusting the frequency of coupler $C$. Substituting $V_b$, $\tilde{g}$ and frequencies of $Q_1$, $Q_2$ and $C$ into Eq. (2), we can obtain direct coupling strength of $Q_1(Q_2)$, $g_{12}/2\pi = 5$ MHz. 
%The coupling strength of $Q_1(Q_2)$ and $C$ is about $g_{1c}(g_{2c})/2\pi = 5$ MHz and two qubits directly coupling strength is about $g_{12}/2\pi = 100$ MHz. 
\begin{figure}[t]
	\includegraphics[width=0.48\textwidth]{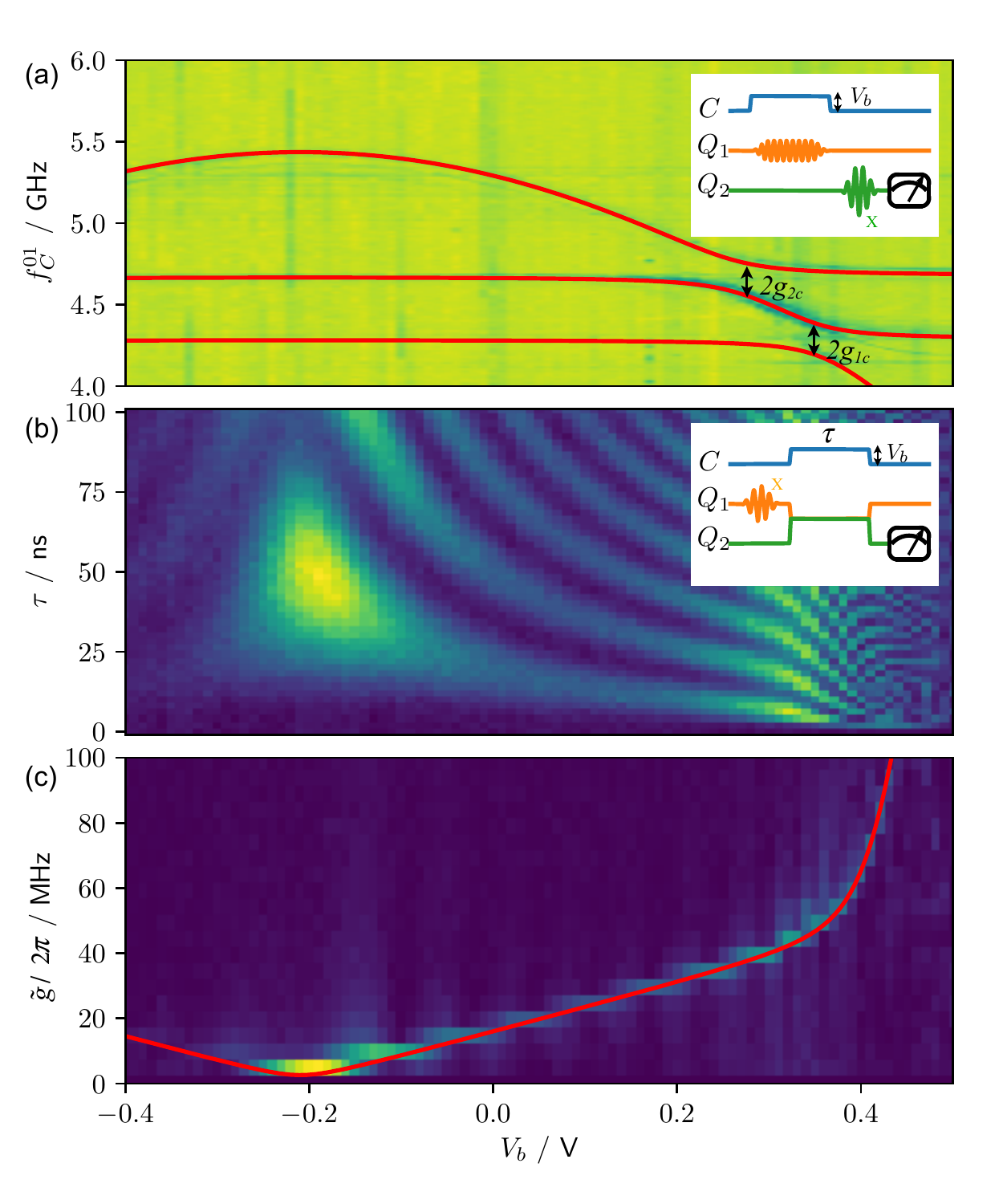}
	\caption{(a) The frequency of the coupler $C$ coupled with qubits $Q_1$ and $Q_2$ versus the amplitude of the bias pulse ($V_b$) of the coupler $C$ and The inset is the schematic diagram. The red curves lines are numerical simulation results fitting the spectrum of qubit-coupler-qubit systems. Two double arrow segments indicate the anti-crossing points and the coupling strength $g_{1c}/2\pi = g_{2c}/2\pi = 100 \, \rm{MHz}$.
	%Coulper $C$ can be Driven by a microwave pulse through qubit $Q_1$ XY control line while biased the coupler by a square pulse. Then readout the excited population of qubit $Q_2$ after A $X_{\pi}$ pulse on qubit $Q_2$.
	(b) The $i$SWAP oscillation when changing the amplitude $V_{b}$ of the bias pulse of the coupler $C$. The inset is the schematic of $i$SWAP experiment. 
	%The $Q_1$ is excited, then tune $Q_2$ resonate with $Q_1$. Biasing the flux of $C$ can change the interaction strength between $Q_1$ and $Q_1$.
	(c) Fourier transform of $i$SWAP oscillation in (b).  The effective coupling strength $\tilde{g}/2\pi$ can vary from 0.40 MHz to 80 MHz. The red curve line shows the calculated results with $g_{12}/2\pi = 5 \, \rm{MHz}$.}
\end{figure}

To easily understand how to operate the qubit-coupler-qubit systems, we use $|Q_{1},C,Q_{2}\rangle$ to describe the energy-eigenstates. When performing CZ gates, the computational state $|101\rangle$ should be an adiabatic evolution from the idle points close to the region where $|101\rangle$ has coupling with another non-computational state (such as $|011\rangle$ or $|200\rangle$)\cite{yan18}, then back to the idling points with no leakage to these non-computational states. Then the state $|101\rangle$ will accumulate a conditional phase. In order to satisfy adiabatic evolution, the gate time should be greater than $1/g_{1c}$ to make sure that there is no leakage to another non-computational state. In our experiment, we firstly bias the $Q_1$, $Q_2$ and $C$ at the frequency of $f_{Q_{1}}^{01} = 4.283$ \rm{GHz}, $f_{Q_{2}}^{01} = 4.679$ \rm{GHz}, and $f_{c}^{01} = 5.419$ \rm{GHz}, respectively, as idling points where the effective ZZ interaction coupling is smaller than 500 KHz and two qubits have better coherence time (the $T_1$ time is  $20.9$ \rm{$\mu$s} for $Q_1$, and $28.8$ \rm{$\mu$s} for $Q_2$). The average single-qubit gate fidelity of $Q_1$ and $Q_2$ are $99.6\%$ and $99.7\%$. Since $Q_1$ and $Q_2$ have stronger coupling strength with $C$, a half-period cosine shape pulse with duration time of 30 ns will satisfy adiabatic evolution condition. 
\begin{figure}[t]
	\includegraphics[width=0.5\textwidth]{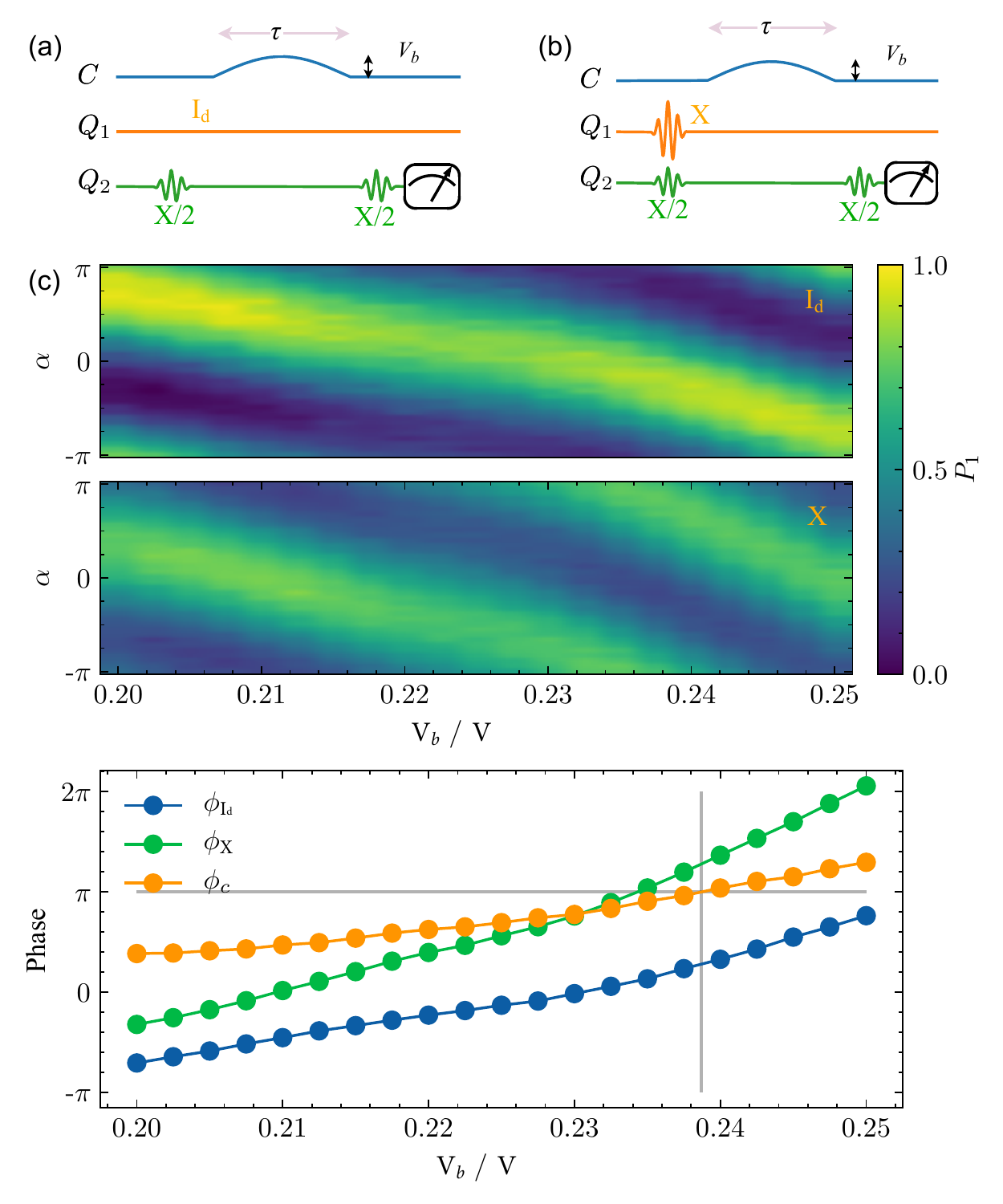}
	\caption{(a) and (b) Schematic of a Ramsey-type experiment measuring the conditional phase of CZ gates . Qubit $Q_1$ is prepared in its ground or excited state. Two $\pi/2$-pulses (the second $\pi/2$-pulses with the phase shift $\alpha$) are on qubit $Q_2$ with an interleaved a idle gate duration of $\tau$ while the CZ gate (half-period cosine shape pulse) on coupler $C$. (c) The oscillations of the population of qubit Q2 are measured by sweeping the amplitude $V_b$ of CZ gate and the second $X/2$ gate phase shift $\alpha$. The top and bottom figures are $Q_1$ in its ground state and its excited state respectively. (d) The initial phases $\phi_{Id}$ and $\phi_{X}$ of the oscillation traces in (c) are fitted as a function of $V_b$, which are displayed by the blue and green lines respectively. The conditional phase $\phi_c = \phi_{X} - \phi_{Id}$ can be extracted as a function of $V_b$(the yellow line). The gray criss-cross line indicates $\phi_c = \pi$ with a proper $V_b$.}
\end{figure}

\begin{figure}[t]
	\includegraphics[width=0.5\textwidth]{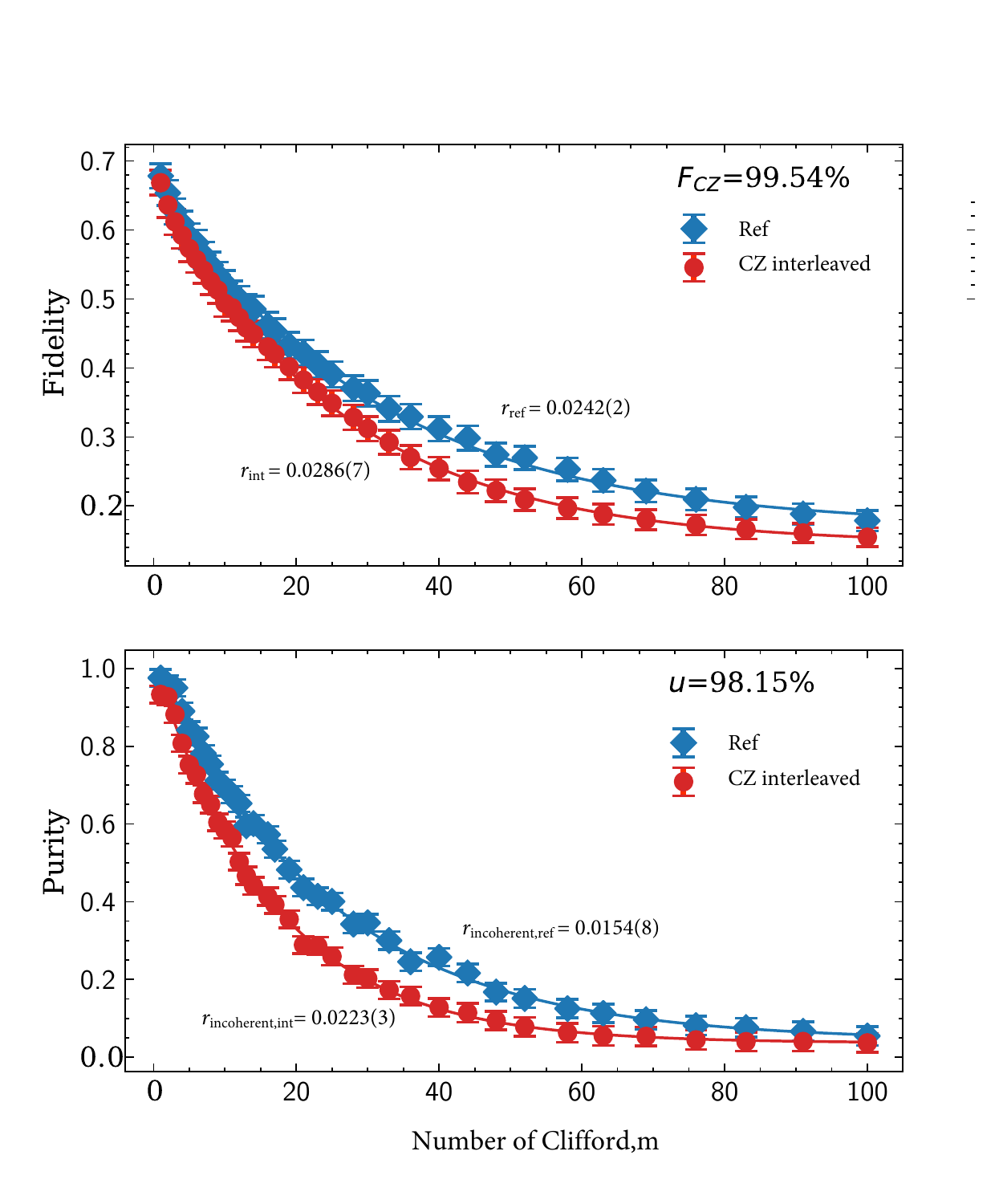}
	\caption{Fidelity and error analysis of the adiabatic CZ gate. (a) Measured sequence fidelity (100 averages) as a function of the number of Cliffords $m$ for both the reference (blue) and interleaved (red) RB experiments. Error bars are the standard deviations from the mean. (b) Measured sequence purity (100 averages) as a function of the number of Cliffords $m$ for both the reference (green) and interleaved (yellow) PB experiments.}
\end{figure}
To calibrate the conditional phase, the Ramsey-type experiments are performed as shown in Fig. 3. Figures 3(a) and (b) are the control pulse sequence and the qubit $Q_1$ is prepared in the ground state $|0\rangle$ and excited state $|1\rangle$, respectively. Figure 3(c) shows the Ramsey oscillation results of qubit $Q_2$ versus the second X/2 gate phase degree $\alpha$ with the different amplitude $V_b$ of coulper $C$ flux pulse. The top and bottom panels are corresponding to $Q_1$ initialized in the ground and excited states, respectively.
So the initial phase $\phi_{Id}$ and $\phi_{X}$ can be extracted by fitting Ramsey oscillation trace. Then we can get the relationship between the conditional phase $\phi_c$ and the $V_b$. As shown in Fig. 3(c), the yellow line is $\phi_c$ versus the $V_b$. We can use $V_b = 0.2382$ V to get the conditional phase $\phi_c = \pi$.
% The final step is to calibrate the single qubit phase to compensate the effect of adiabatic CZ gate\cite{mcka2017} using the quantum process tomography.
 $\phi_{Id}$ is a local dynamical phase accumulated on $Q_2$ which can be easily compensated by a virtual $Z$ gate \cite{mcka2017}. Exchange the roles of $Q_1$ and $Q_2$ in the same experiment, local phase on $Q_1$ can be measured too.

The Clifford based Randomized Benchmarking (RB) method can be used to assess and optimize the performance of adiabatic CZ gates\cite{bare14,Knill08,Kelly2014,Magesan11,Magesan12,Malley15}, which is performed by applying a gate sequence of $m$ two-qubit Clifford gates followed by an additional $\big(m + 1\big)$th gate to invert the whole sequence. Fig. 4(a) shows the sequence fidelity as a function of $m$ for both the reference and CZ-interleaved cases. The results are the average of 100 random samples. The parameters of adiabatic CZ gate are obtained by implementing Nelder-Mead (NM) optimization algorithm in 100 evaluations. We can obtain the decay constants $p_{\rm{ref}}$ and $p_{\rm{int}}$ from the exponential function $F = Ap^{m} + B$ fitting, and then the error rate of per Clifford gate $r_{\rm{ref}}$ and $r_{\rm{int}}$ from $r = 3/4 (1-p)$. The CZ gate error rate can be extracted from $r_{\rm{CZ}} = 3/4(1-p_{\rm{int}}/p_{\rm{ref}})$ and the CZ gate fidelity from $F_{\rm{CZ}} = 1 - r_{\rm{CZ}}$. 
The average error rate of per Clifford gate of our scheme is $r_{\rm{ref}}= 0.0242(2)$, and $r_{\rm{int}}= 0.0286(7)$ for interleaved, as shown in Fig. 4(a). The CZ gate error $r_{\rm{CZ}} = 0.0046(2)$. The fidelity of adiabatic CZ gate is $F_{\rm{\rm{CZ}}} = 99.53(8)\%$. To estimate the decoherence error, we also measure the two-qubit purity benchmarking (PB)\cite{wall2015,feng2016} as shown in Fig. 4(b). We also use exponential function $F = A^{'}u^{m-1} + B^{'}$ to fit sequence purity as a function of $m$, and obtain the purity error rate from  $r_{\rm{incoherent}} = 3/4(1-\sqrt{u})$. The average purity error of per Clifford gate is $r_{\rm{incoherent, int}} = 0.0223(3)$. The average purity error of interleaved sequence is $r_{\rm{incoherent, ref}} = 0.0154(8)$. 
We estimate incoherence error contribution is $r_{\rm{incoherent, ref}}/r_{\rm{ref}} = 64\%$ of total errors per two-qubit Clifford gate. 
%So in order to improve the gate fidelity, we need to increase the coherence times or decease the duration time of CZ gate.

\begin{figure}[t]
	\includegraphics[width=0.5\textwidth]{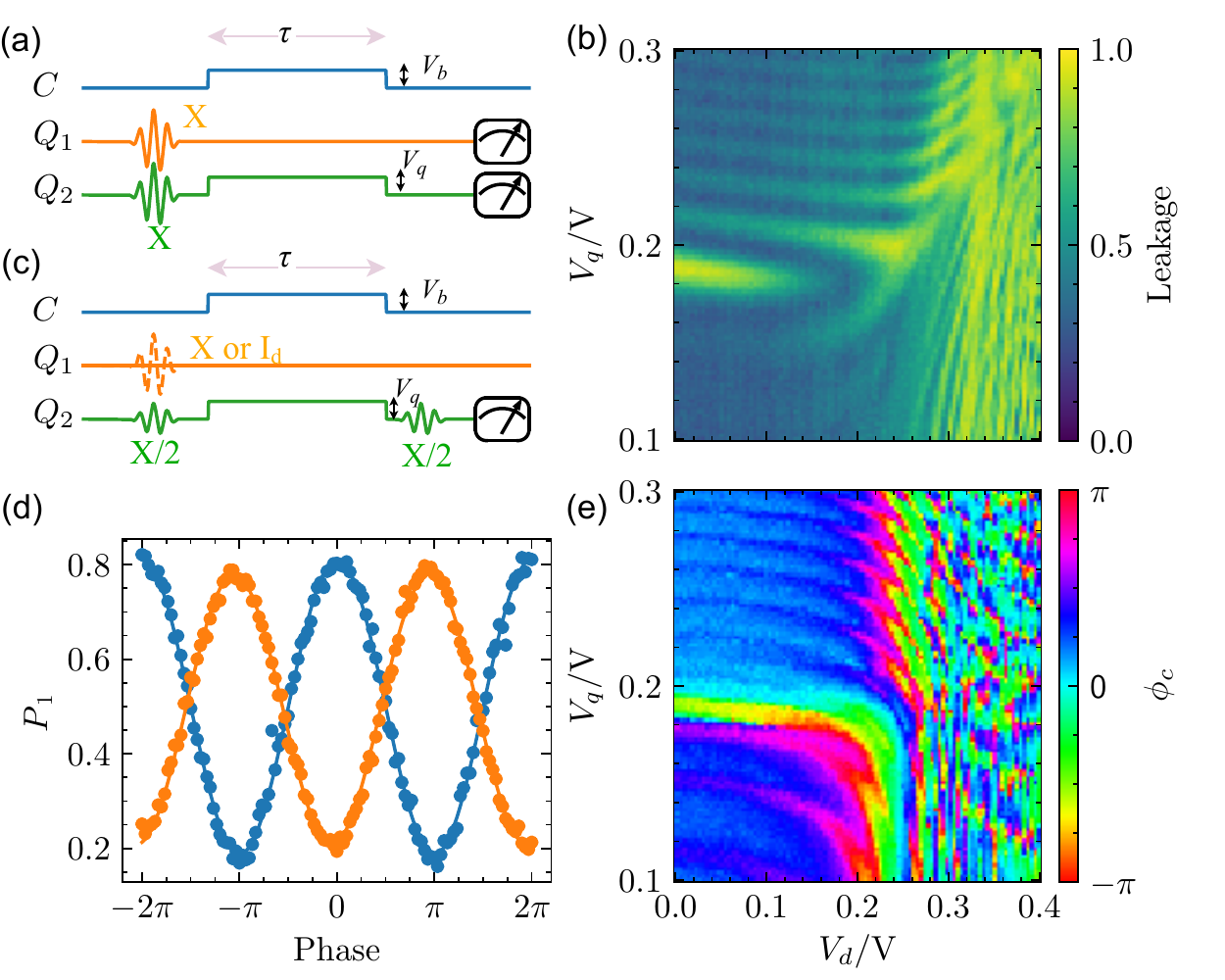}
	\caption{(b) The control sequence to measure leakage errors when implentmenting the diabatic CZ gates. (b) Leakage error results after CZ gates. (c) Schematic of a Ramsey-type experiment measuring the conditional phase of diabatic CZ gates. The pulse shape of diabatic CZ gates is a square-shaped pulse of duration $\tau = 18$ ns. Qubit $Q_1$ is prepared in its ground or excited state. (d) The results of the Ramsey oscillations under $V_b = 0.173 \, \rm{V}$ and $V_q = 0.174 \, \rm{V}$. The conditional phase $\phi_c = \pi$.  (e) The conditional phase results after CZ gates.}
\end{figure}
\begin{figure}[t]
	\includegraphics[width=0.5\textwidth]{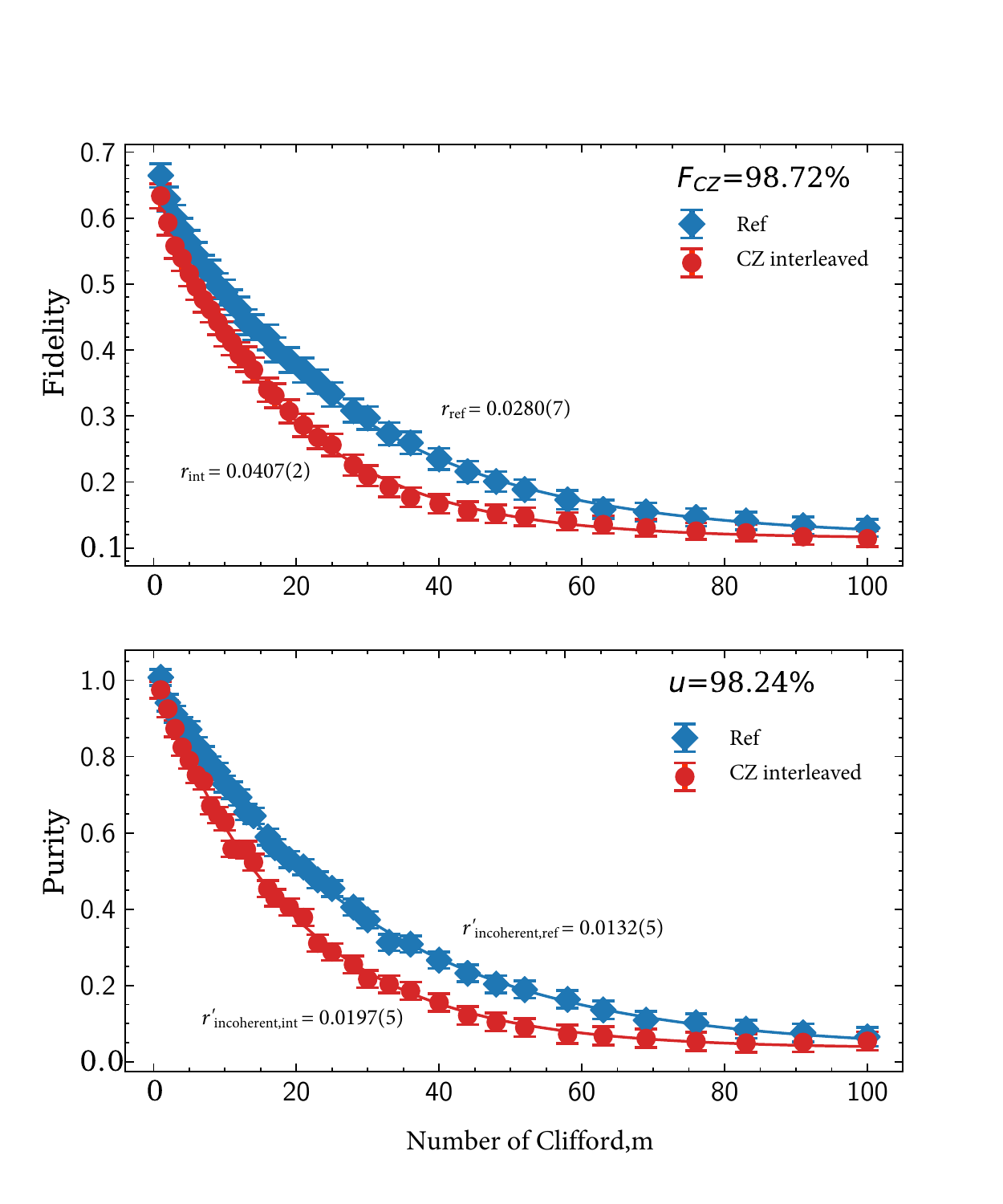}
	\caption{Fidelity and error analysis of the diabatic CZ gate. (a) Measured sequence fidelity (100 averages) as a function of the number of Cliffords $m$ for both the reference (blue) and interleaved (red) RB experiments. Error bars are the standard deviations from the mean. (b) Measured sequence purity (100 averages) as a function of the number of Cliffords $m$ for both the reference (green) and interleaved (yellow) PB experiments.}
\end{figure}
To realize a faster CZ gate, the two-qubit gates need to have a stronger effective coupling strength $\tilde{g}$. From Fig. 2(c), the maximum of $\tilde{g}$ is up to $80 $ MHz. So we can choose $\tau \approx 18$ ns as the duration time of CZ gate. However, the evolution of CZ gate will not satisfy adiabatic conditions, which will induce the leakage error. To simplify diabatic CZ gate operation, the square-shaped pulse is used with distortion corrected\cite{Rol19}. So we have two main controllable parameters, the coupling $\tilde{g}$ and frequency detuning $\Delta = f_{Q_{2}}^{01}-f_{Q_{1}}^{01}$ to realize the diabatic CZ gates. To adjust $\Delta$ value in experiments, we fix $Q_1$ frequency at the idle points and change the flux bias pulse amplitude $V_q$ of $Q_2$. Before calibrating the conditional phase, we should exactly identify the leakage versus coupling strength $\tilde{g}$ and frequency detuning $\Delta$. As shown in Fig. 5(a), we initialize two qubits in the excited state and then perform the diabatic CZ gate. If there have the leakage occurs, the ground state population of lower frequency qubit will increase. By sweeping the voltage of $V_b$ and $V_q$, we can map out leakage errors. The result is shown in Fig. 5(b), where the dark region means the low-leakage errors. Then to calibrate the conditional phase by executing Ramsey-type experiments and we can get the relationships between $\phi_c$ and $V_b$, $V_q$. Figure 5(c) shows the experimental control pulse sequence which is similar to the adiabatic CZ gates except that the square-shaped pulse  is applied and the values of $Q_2$ flux bias pulse amplitude $V_q$ are explored. The measurement result is shown in Fig. 5(e). The red regions are $\phi_c$ close to $\pi$ or $-\pi$. So we can pick up $V_b$ and $V_q$ values to satisfy $\phi_c = \pi$, then check leakage errors in Fig. 5(b). We find $V_b = 0.173 \, \rm{V}$ and $V_q = 0.174 \, \rm{V}$ satisfying both constraint conditions. Figure 5(d) shows the results of the qubit $Q_2$ Ramsey oscillations under these conditions. The phase difference of two oscillations is $\pi$, which means the conditional phase $\phi_c = \pi$.

Then RB is used to optimize the fidelity of CZ gate by NM algorithm. The best optimization result is shown in Fig. 6(a). The optimized CZ gate time is 19 ns. The average error rate of per Clifford gate $r'_{\rm{ref}}= 0.0280(7)$ and $r'_{\rm{int}}= 0.0407(2)$. The CZ gate error rate $r'_{\rm{CZ}} = 0.0127(8)$. The fidelity of diabatic CZ gate is $F_{\rm{\rm{CZ}}} = 98.72(2)\%$. As shown in Fig. 6(b), The sequence purity $r'_{\rm{incoherent,int}}$ and $r'_{\rm{incoherent,ref}}$ are 0.0197(5) and 0.0132(5), respectively. The incoherence error contributes $47\%$ of total errors per two-qubit Clifford gate.  
%The incoherent error is $46.2\%$ of total errors. Other errors arise from the non-perfect quantum state control.

In summary, we experimentally implemented a tunable-coupled two qubits system, and
realized high fidelity CZ gate in two different schemes. The highest RB fidelity of the adiabatic and diabatic
CZ gates we achieved were $99.53(8)\%$ and $98.72(2)\%$, respectively. 
The adiabatic CZ gate defined by a cosine
flux pulse on coupler requiring fewer calibrations compared to diabatic CZ gate,
which was implemented by applying parallel square-sharped flux pulses both on coupler and one frequency
tunable qubit. Purity benchmarking which discribe incoherence error was also stutied. 
We found that incoherence error contributed $64\%$ and $47\%$ in total error per Clifford
gate for adiabatic or diabatic CZ gates, respectively.

This work was supported by the NSFC of China (Grants No. 11890704, 12004042, 11674376), the NSF of Beijing (Grants No. Z190012), National Key Research and Development Program of China (Grant No. 2016YFA0301800) and the Key-Area Research and Development Program of Guang-Dong Province (Grants No. 2018B030326001)..

%\newpage

\end{document}